\documentclass[twocolumn,twoside,preprintnumbers,amsmath,amssymb,pacs]{revtex4}
\usepackage{epsfig}
\usepackage{graphicx}
\usepackage{dcolumn}
\usepackage{bm}

\usepackage{fancyhdr}

\usepackage{pslatex}

\usepackage{psfrag}

\begin{document}

\title{{\Large Exploring the infrared gluon and ghost propagators using large asymmetric lattices}}

\author{O. Oliveira and P. J. Silva}

\affiliation{ Centro de F\'{\i}sica Computacional \\
Departamento de F\'{\i}sica, Universidade de Coimbra \\
P-3004-516 Coimbra\\
Portugal }


\received{on 20 October, 2006}

\begin{abstract}
We report on the infrared limit of the
quenched lattice Landau gauge gluon propagator computed from large
asymmetric lattices. In particular, the compatibility of the pure power
law infrared solution $( q^2 )^{2\kappa}$ of the Dyson-Schwinger equations
is investigated and the exponent $\kappa$ is measured. The lattice data
favours $\kappa \sim 0.52$, which would imply a
vanishing zero momentum gluon propagator as predicted by the Kugo-Ojima
confinement mechanism and the Zwanziger horizon condition. 
Results for the ghost propagator and for the running coupling constant
are shown.


\end{abstract}

\maketitle

\thispagestyle{fancy}
\setcounter{page}{1}

\section{Introduction and motivation}

The mechanism of quark and gluon confinement is not fully understood yet. 
The study of the fundamental Green's functions of Quantum Chromodynamics 
(QCD), namely the 
gluon and ghost propagators, can help in the understanding of such 
mechanism. Indeed, there are gluon confinement criteria connected with the 
behaviour of the propagators at zero momentum. In particular, the Zwanziger 
horizon condition implies a null zero momentum gluon propagator $D(q^2)$,
 and the 
Kugo-Ojima confinement mechanism requires an infinite zero momentum ghost 
propagator $G(q^2)$. The violation of positivity for the gluon propagator 
can also be seen as a signal for confinement \footnote{For details on these 
topics see, for example, \cite{AlkLvS01} and references therein.}.

In QCD, the investigation of its infrared limit has to rely
 on non-perturbative methods like 
the Dyson-Schwinger equations (DSE) and lattice QCD. Both methods have good 
and bad features, namely we can solve analitically the DSE in the infrared,
 but one has to rely on a truncation of an infinite tower of equations. On 
the lattice, one includes all non-perturbative physics,
 but one has to care about finite volume and finite lattice spacing effects.

Recent studies of Dyson-Schwinger equations obtained a pure power law 
behaviour for the infrared gluon and ghost dressing functions,

\begin{eqnarray}
Z_{gluon}(q^2)\equiv q^2 D(q^2) \sim(q^2)^{2\kappa}\\
Z_{ghost}(q^2)\equiv q^2 G(q^2) \sim(q^2)^{-\kappa},
\end{eqnarray}

\noindent
with $\kappa=0.595$ \cite{lerche}, which implies a vanishing (infinite) gluon 
(ghost) propagator for zero momentum. 
Studies using functional renormalization equations  \cite{pawl,gies} provided
 bounds in the possible 
values of the infrared exponent $0.52 < \kappa < 0.595$. The  
infrared exponent obtained from time independent stochastic quantisation 
\cite{tisq} is within these bounds ($\kappa=0.52145$).

\begin{figure}[htbp]
\begin{center}
\vspace*{0.7cm}
\psfrag{EIXOX}{{\small $q(GeV)$}}
\psfrag{EIXOY}{{\small $Z_{gluon}(q^2)$}}
\psfrag{GEIXY}{{\small $Z_{ghost}(q^2)$}}
\includegraphics[width=8cm]{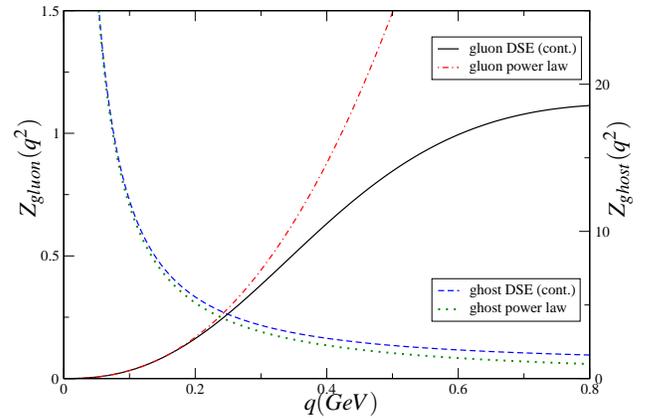}
\caption{ The DSE solution for the gluon and ghost dressing functions 
\cite{fish.penn} compared with the corresponding pure power laws.}
\label{dsepower}
\end{center}
\end{figure}

As an infrared (IR) analytical solution of DSE, the pure power law is valid 
only for very low momenta. In figure \ref{dsepower}, the DSE gluon and 
ghost propagators 
\cite{fish.penn} are compared with the corresponding pure power law 
solution. Note that for the gluon propagator the power law is valid 
only for momenta below 200 MeV, and for the ghost the infrared solution is 
restricted to even lower momenta. If one wants to use 
lattice QCD to check for a pure power law behaviour, certainly one should 
consider a lattice volume sufficiently large to accomodate a minimum 
number of points in the region of interest. For symmetric lattices this would 
require a too large volume. A cheaper solution is the use of large 
asymmetric lattices $L_s^3 \times L_t$, with $L_t \gg L_s$. For example, 
in \cite{sardenha, dublin, finvol, nossoprd, madrid, tucson} we use a set 
of asymmetric lattices with $L_t=256$ and $L_s=8,10,\ldots,18$ 
(see \cite{nossoprd} for the technical details of the simulations).  
The large temporal size of these lattices allow to access to momenta 
as low as 48 MeV. Of course, the price to pay 
are finite volume effects caused by the small value of $L_s$.

Recently, some criticism have been raised against the use of asymmetric 
lattices to study the infrared limit of QCD Green's functions. In 
\cite{cucc06}, the authors studied the gluon and ghost propagators in SU(2) 
3-d theory and reported systematic effects due to the 
 asymmetry of the lattice, if one compares with a symmetric one. Of course, 
this comparison cannot be done in our case. Also in \cite{phdstern} the 
author reported asymmetry effects in the propagators. However, in what 
concerns the gluon propagator, such effects have been already investigated 
in \cite{finvol}. Furthermore, in \cite{finvol,nossoprd} it is shown that 
the approach $L_s \to +\infty$ is smooth. Indeed, a quick look at the gluon 
and ghost data in \cite{cucc06}, again, suggests that the approach 
$L_s \to +\infty$ is smooth, i. e. that the largest symmetric propagators 
can be obtained by extrapolation of the asymmetric ones.
In what concerns the quantitative results coming from a single large
asymmetric
lattice, all the above studies show that they provide, at least, a bound
on the infinite volume limit. 
 Here we report on 
the status of our investigations concerning the use of asymmetric lattices 
to study the infrared properties of QCD.
In what concerns the positivity violation 
of the gluon propagator, our results can be seen in \cite{tucson}.

\section{Extracting the infrared exponent $\kappa$ from the gluon propagator}

\begin{figure}[htbp]
\begin{center}
\vspace*{0.7cm}
\psfrag{EIXOX}{{$q(GeV)$}}
\psfrag{EIXOY}{{$D(q^2)$}}
\includegraphics[width=8cm]{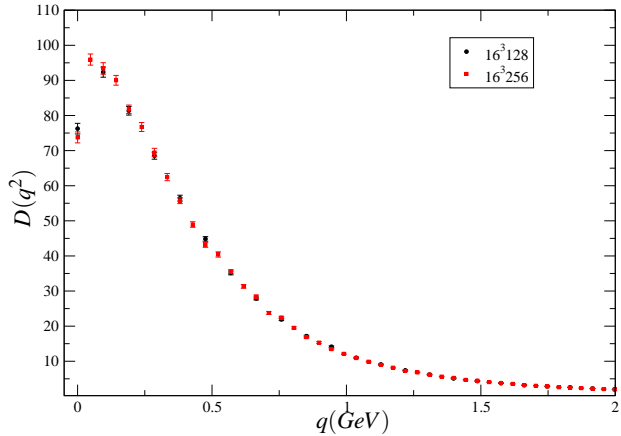}
\caption{The gluon propagator for $16^3\times 128$ and $16^3\times 256$ lattices.}
\label{ls16}
\end{center}
\end{figure}

In \cite{nossoprd}, we have computed the gluon propagator 

\begin{equation}
D^{ab}_{\mu\nu}(q) = \delta^{ab} \Big(  \delta_{\mu\nu} - \frac{q_\mu 
q_\nu}{q^2} \Big) D( q^2 )
\end{equation}

\noindent
for SU(3) four-dimensional asymmetric lattices 
$L_s^3 \times 256$, with $L_s=8,10,\ldots,18$. To check that 
the temporal size is large enough, 164 $16^3 \times 128$ Wilson action 
gauge configurations were generated. In what concerns the gluon propagator 
for time-like momenta, there are no differences between the $16^3\times 128$ 
and $16^3\times 256$ data (see figure \ref{ls16}). This gives us 
confidence that the temporal extension of our lattices is sufficiently large.

\begin{figure}[htbp]
\begin{center}
\vspace*{0.3cm}
\includegraphics[width=8cm]{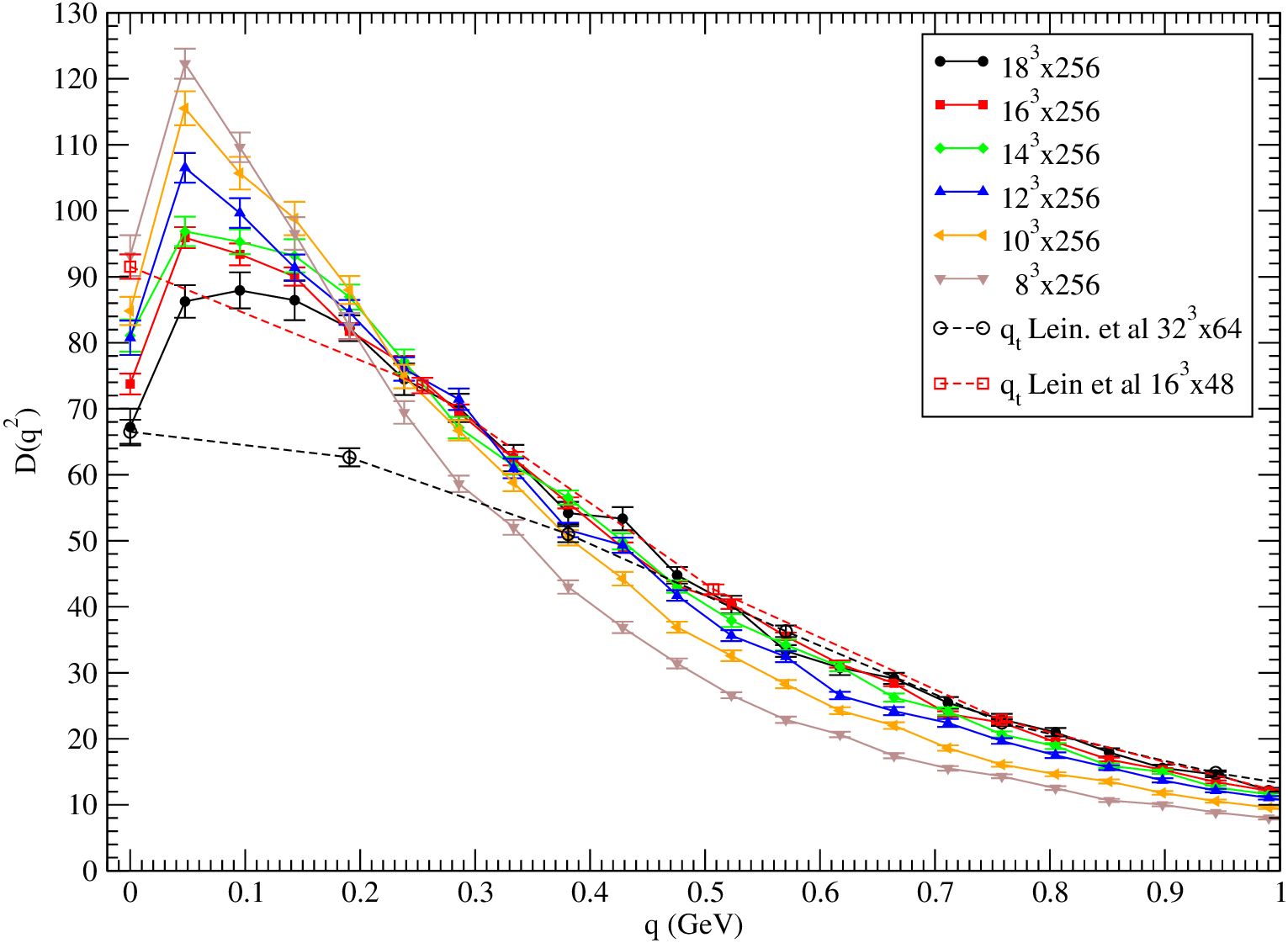}
\caption{The gluon propagator for all lattices $L_{s}^{3}\times256$, 
considering 
only pure temporal momenta. For comparisation, we also show the 
$16^3\times 48$ and $32^3\times 64$ propagators computed in \cite{lein99}.}
\label{finvolir}
\end{center}
\end{figure}

In what concerns the spatial size, it was observed that the propagator 
depends on the spatial size of the lattice (see figure \ref{finvolir}). 
The gluon propagator decreases with the volume for the smallest momenta 
and increases with the volume for higher momenta.
 
In order to compute the infrared exponent $\kappa$ from the 
lattice data, we considered fits of the smallest temporal momenta of the 
gluon dressing function $Z_{gluon}(q^2)$ to 
a pure power law with and without polinomial corrections. The results 
are in table 
\ref{kappatab}. In general, the $\kappa$ values increase with the volume 
of the lattice. So, our $\kappa$ can be read as lower bounds in the 
infinite volume figure $\kappa_{\hspace{1mm}\infty}$.

\begin{table}[htbp]
\begin{center}
\small
\begin{tabular}{|c|@{\hspace{0.3cm}}r@{\hspace{0.3cm}}r|
                                     r@{\hspace{0.3cm}}r|
                                     l@{\hspace{0.3cm}}r|}
\hline
   $L_s$     & \multicolumn{2}{c|}{$z (q^2)^{2 \kappa}$} 
               & \multicolumn{2}{c|}{$z (q^2)^{2 \kappa} ( 1 + a q^2 )$} 
               & \multicolumn{2}{c|}{$z (q^2)^{2 \kappa} ( 1 + a q^2 + b q^4)$} \\
\hline
                      &                      &     & & & & \\
   $ 8$  & $0.4496^{+22}_{-29}$   & 2.14 &
                        $0.4773^{+37}_{-52}$   & 0.02 &
                        $0.4827^{+75}_{-74}$   & 0.00 \\
                      &                      &     & & & & \\
   $10$  & $0.4650^{+31}_{-37}$   & 0.10 &
                        $0.4827^{+49}_{-68}$   & 0.25 &
                        $0.4765^{+104}_{-99}$  & 0.14 \\
                      &                      &     & & & & \\
   $12$  & $0.4663^{+30}_{-36}$   & 1.19 &
                        $0.4822^{+51}_{-69}$   & 0.21 &
                        $0.4849^{+94}_{-97}$   & 0.18 \\
                      &                      &     & & & & \\
   $14$  & $0.4918^{+26}_{-40}$   & 0.09 &
                        $0.5053^{+52}_{-67}$   & 0.16 &
                        $0.4992^{+100}_{-80}$   & 0.06 \\
                      &                      &     & & & & \\
   $16$  & $0.4859^{+22}_{-24}$   & 0.40 & 
                        $0.5070^{+36}_{-50}$   & 0.44 & 
                        $0.5131^{+67}_{-64}$   & 1.03  \\
                      &                      &     & & & & \\
   $18$  & $0.5017^{+49}_{-40}$   & 0.20 & 
                        $0.5169^{+89}_{-70}$   & 0.00 &
                        $0.514^{+12}_{-15}$    & 0.00     \\
                      &                      &     & & & & \\
\hline
\end{tabular}
\caption{$\kappa$ and $\chi^2/d.o.f.$ from fitting the gluon  
dressing function computed from the different lattices $L_{s}^{3}\times256$.
In all fits, the range of momenta was chosen such that the fits have one 
degree of freedom. The lowest momentum considered being always the first 
nonvanishing momentum.
The errors shown are statistical and were computed 
using the bootstrap method, with the number of bootstrap samples being about 
ten times the number of configurations. }
\label{kappatab}
\end{center}
\end{table}

Considering a linear or quadratic dependence on the inverse of the volume for 
the infrared exponent, we can try to extrapolate the figures of table 
\ref{kappatab}. In what concerns the results for the pure power law fits, they 
are not described by these functional forms. Using the 
corrections to the pure power law, one gets values for  
$\kappa_{\infty}$ in the interval 
$[0.51,0.56]$; the weighted mean value being 
$\bar{\kappa}_{\infty}= 0.5246(46)$.

\begin{figure}[htbp]
\begin{center}
\vspace*{0.7cm}
\psfrag{EIXOX}{{\Large $q(GeV)$}}
\psfrag{EIXOY}{{\Large $D(q^2)$}}
\includegraphics[width=8cm]{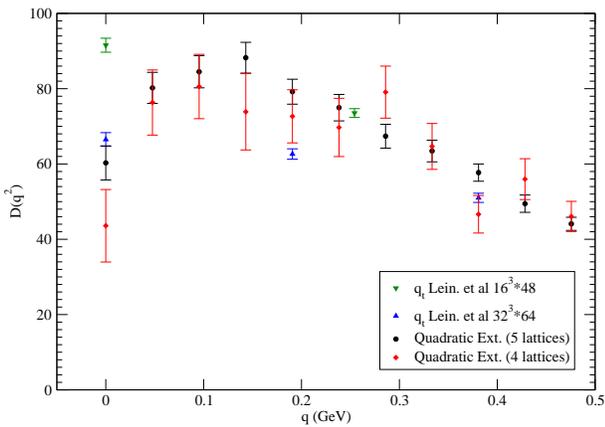}
\end{center}
\caption{The bare extrapolated gluon propagator for momenta below $500MeV$. 
The errors on the propagator were computed assuming a gaussian error 
propagation.}
\label{gpextzir}
\end{figure}

On the other hand, one can also extrapolate directly the gluon propagator, 
as a function of the inverse of the volume,
to the infinite volume limit, fitting each timelike momentum propagator 
separately. Doing so, we are 
taking the limit $L_s \to \infty$ , assuming a sufficient number of points 
in the temporal direction. Several types of polinomial extrapolations were 
tried, using different sets of lattices, and we conclude that the data is 
better described by quadratic extrapolations of the data from the 
4th and 5th largest lattices. 
In figure \ref{gpextzir} the two extrapolations of the gluon propagator 
are shown. For comparisation, we also include the $16^3\times 48$ and 
$32^3\times 64$ propagators computed in \cite{lein99}.

The values of $\kappa$ extracted from these extrapolated propagators are 
$\kappa = 0.5215(29)$, with a $\chi^2/d.o.f. = 0.02$, using the largest 
5 lattices in the extrapolation, and $\kappa = 0.4979(66)$, 
$\chi^2 / d.o.f. = 0.27$ using the largest 4 lattices. Fitting the 
extrapolated data to the polinomial corrections to the pure power law, 
one can get higher values for $\kappa$. Note that the first value is on 
the top of the value obtained from extrapolating directly $\kappa$ as a 
function of the volume. In conclusion, one can claim a 
$\kappa \in [0.49,0.53]$, 
with the lattice data favouring the right hand side of the interval.

\section{Are DSE fitting forms reproduced by lattice data?}

In this section we will try to check if the 
pure temporal lattice data of the gluon propagator 
can be described by some functional 
forms that fitted the solution for the gluon propagator in 
Landau gauge of the Dyson-Schwinger equations reported in 
\cite{fischer04, fischer03}. Only the largest lattices will 
be considered. 

\subsection{The infrared region}

In the IR region, the continuum DSE solution is well described by the 
following expressions \cite{fischer04},

\begin{equation}
Z_{cut}(q^2) = \omega 
    \left(\frac{q^2}{\Lambda_{QCD}^2+q^2}\right)^{2\kappa}\, ,
\end{equation}

\begin{equation}
  Z_{pole}(q^2) = \omega
  \frac{\left(q^2\right)^{2\kappa}}
       {\left(\Lambda_{QCD}^2\right)^{2\kappa}+\left(q^2\right)^{2\kappa}}\, .
\end{equation}

\noindent
The first functional form has a branch cut in the IR region. The last one has
a pole in the IR region.

\begin{figure}[htbp]
\begin{center}
\psfrag{EIXOX}{$q(GeV)$}
\psfrag{EIXOY}{$q^2 D(q^2)$}
\includegraphics[width=8cm]{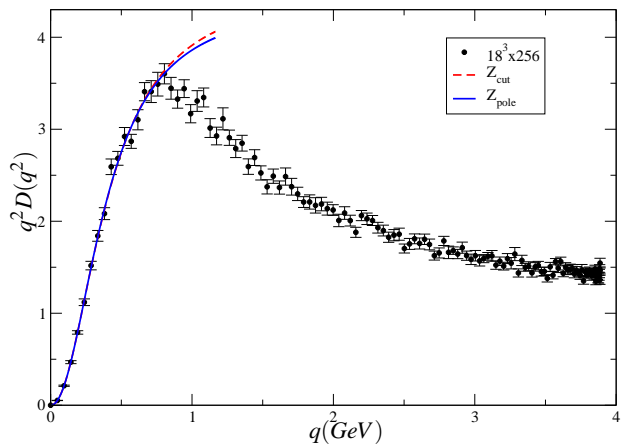}
\caption{The lattice gluon dressing function compared with IR fits. }
\label{IRcomp}
\end{center}
\end{figure}

In what concerns the lattice propagator, both formulas provide good fits
up to the maximum of the dressing function --- see table \ref{IRfits} and  
figure \ref{IRcomp}. However, the measured infrared exponents
are larger than those obtained assuming a pure power law for the IR region 
\cite{nossoprd}, and support an infrared vanishing gluon propagator.

\begin{table}[htbp]
\begin{center}
\begin{tabular}{cccccc}
\hline 
\rule[-1mm]{0pt}{5mm} & Lattice & \hspace{0.2cm} $q_{max}$ \hspace{0.2cm}
          & \hspace{0.2cm} $\kappa$ \hspace{0.2cm}
          & $\Lambda_{QCD}$
          & $\chi^2 /d.o.f.$  \\
\hline
\rule{0pt}{4mm} $Z_{cut}$ & $16^3\times 128$ & $570$ 
                              & $0.5117^{+48}_{-46}$ 
                              & $417^{+8}_{-8}$
                              &  $1.25$ \\
\rule{0pt}{4mm}           & $16^3\times 256$ & $ 664$ 
                              & $0.5090^{+19}_{-20}$ 
                              & $409^{+4}_{-4}$
                              &  $0.71$ \\
\rule[-1mm]{0pt}{5mm}           & $18^3\times 256$ & $711$ 
                              & $0.5320^{+28}_{-30}$ 
                              & $389^{+5}_{-6}$
                              &  $1.14$ \\
\hline
\rule{0pt}{4mm} $Z_{pole}$ & $16^3\times 128$ & $570$ 
                              & $0.5100^{+38}_{-31}$ 
                              & $416^{+6}_{-8}$
                              &  $1.15$ \\
\rule{0pt}{4mm}           & $16^3\times 256$ & $664$ 
                              & $0.5077^{+16}_{-17}$ 
                              & $409^{+4}_{-3}$ 
                              &  $0.69$ \\
\rule[-1mm]{0pt}{5mm}           & $18^3\times 256$ & $711$ 
                              & $0.5266^{+29}_{-21}$ 
                              & $391^{+3}_{-7}$ 
                              &  $1.09$ \\
\hline
\end{tabular}
\caption{IR fits. $q_{max}$ is the highest momenta considered in the fit. 
$q_{max}$ and $\Lambda_{QCD}$ are given in MeV.}
\label{IRfits}
\end{center}
\end{table}

\vspace*{0.3cm}

\subsection{The full range of momenta}

The lattice propagator, for the full range of pure temporal momenta, was fitted
to
  
\begin{equation}
   Z_{fit}(q^2) ~ = ~ B(q^2) \, 
                      \left( \alpha ( q^2 ) \right)^{-\gamma}\, , 
               \hspace{0.2cm}  \gamma = -13/22\, ,
\end{equation}

\noindent
with $B(q^2)=Z_{pole,cut}$. Two different definitions for the running coupling
were considered \cite{fischer04,fischer03}:

\begin{eqnarray}
  \alpha_{P}(q^2) & = &
       \frac{1}{1+\frac{q^2}{\Lambda_{QCD}^2}}
       \bigg[\alpha(0)+(q^2/\Lambda_{QCD}^2)\times \nonumber \\
  & &  \frac{4\pi}{\beta_0}
       \Big(\frac{1}{\ln(q^2/\Lambda_{QCD}^2)}-\frac{1}{q^2/\Lambda_{QCD}^2-1}
       \Big)\bigg]\, ;
\\
  \alpha_{LN}(q^2) & = & 
       \frac{\alpha(0)}{\ln\left[e+a_1(q^2/\Lambda_{QCD}^2)^{a_2}\right]}\, .
\end{eqnarray}

\begin{figure}[htbp]
\begin{center}
\vspace*{0.7cm}
\psfrag{EIXOX}{{\small $q(GeV)$}}
\psfrag{EIXOY}{{\small $q^2 D(q^2)$}}
\includegraphics[width=8cm]{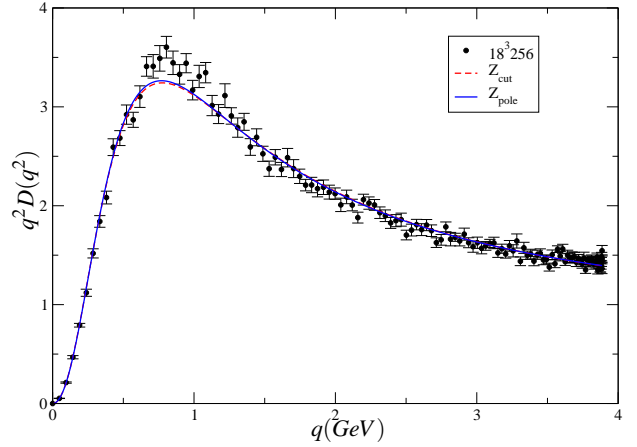}
\caption{Compatibility between our lattice data and global fits using $\alpha_{P}(q^2)$.}
\label{comp.18.alpha1}
\end{center}
\end{figure}

The fits using $\alpha_{\hspace{1mm}P}(q^2)$ (using $\beta_0=11$) for the running coupling
have a $\chi^2/d.o.f. \ge 2$ for $L_s = 16$. However, for the largest lattice,
$18^3 \times 256$, the data is well described; see table \ref{fit.all.alpha.p} 
and figure \ref{comp.18.alpha1}.

\begin{table}[htbp]
\begin{center}
\begin{tabular}{lccccc}
\hline
\rule[-1mm]{0pt}{5mm}    $18^{3}\times 256$
  & $\alpha(0)$
  & $\kappa$
  & $\Lambda_{QCD}$
  & $\chi^2 /d.o.f.$  \\
\hline
\rule{0pt}{4mm}$B=Z_{cut}$ & $10.92^{+13}_{-18}$ 
                 & $0.5280^{+30}_{-22}$ 
                 & $549^{+2}_{-3}$ 
                 & $1.61$ \\
\rule[-1mm]{0pt}{5mm}$B=Z_{pole}$ & $10.01^{+14}_{-17}$ 
                  & $0.5263^{+19}_{-22}$  
                  & $550^{+2}_{-3}$ 
                  & $1.54$ \\
\hline
\end{tabular}
\caption{Fits to all lattice data using $\alpha_{P}(q^2)$.}
\label{fit.all.alpha.p}
\end{center}
\end{table}

Note that only when one uses $B=Z_{pole}$ the $\kappa$ values agree 
with those from the IR fits. However, $\Lambda_{QCD}$ is not 
compatible with the IR values.
The fitting form differs from the lattice data mainly at the 
maximum of the dressing function ($\sim 0.8$ GeV).

\begin{figure}[t]
\begin{center}
\psfrag{EIXOX}{{\small $q(GeV)$}}
\psfrag{EIXOY}{{\small $q^2 D(q^2)$}}
\includegraphics[width=8cm]{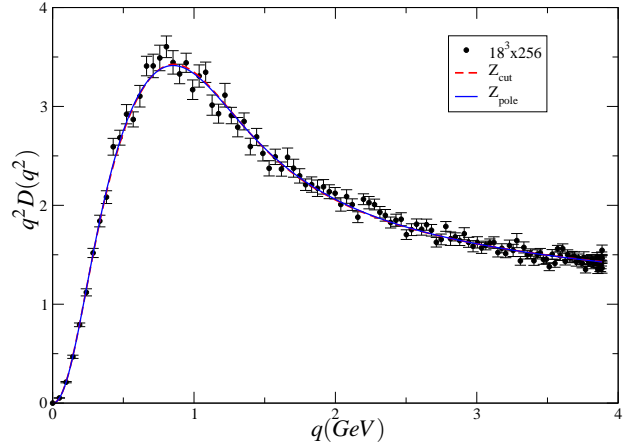}
\caption{Compatibility between our lattice data and global fits using 
$\alpha_{LN}(q^2)$.}
\label{comp.18.alpha2}
\end{center}
\end{figure}

The lattice data adjusts better to $\alpha_{LN}(q^2)$ than to the previously
consi\-de\-red running coupling. Indeed, looking at the fitting results for our
largest lattice (see tables \ref{dse2_b_cut} and \ref{dse2_b_pole}), the 
values of $\kappa$ and $\Lambda_{QCD}$ are 
essentially the values obtained in the IR study. Moreover, on overall there is 
good agreement between the fitting function and the lattice data, see figure 
\ref{comp.18.alpha2}.

\vspace*{-0.2cm}

\begin{table}[h]
\begin{center}
\begin{tabular}{lccccc}
\hline
\rule[-1mm]{0pt}{5mm}    $B=Z_{cut}$
  & $\kappa$
  & $\Lambda_{QCD}$
  & $a_1$
  & $a_2$
  & \hspace{-0.2cm}$\chi^2 /d.o.f.$  \\
\hline
\rule{0pt}{4mm}$16^3\times 128$ & $0.5435^{+36}_{-41}$ 
                 &  $364^{+4}_{-4}$ 
                 & $0.0062^{+3}_{-3}$ 
                 & $2.44^{+2}_{-1}$  
                 & $1.82$ \\
\rule{0pt}{4mm}$16^3\times 256$ & $0.5244^{+21}_{-15}$ 
                 & $374^{+2}_{-2}$  
                 & $0.0072^{+3}_{-3}$ 
                 & $2.42^{+1}_{-1}$ 
                 & $1.73$ \\
\rule[-1mm]{0pt}{5mm}$18^3\times 256$ & $0.5368^{+25}_{-24}$ 
                 & $381^{+3}_{-4}$  
                 & $0.0067^{+4}_{-5}$ 
                 & $2.55^{+2}_{-2}$ 
                 & $1.23$ \\
\hline
\end{tabular}
\caption{Fits to all lattice data using $\alpha_{LN}(q^2)$ and $B=Z_{cut}$.
\label{dse2_b_cut}}
\end{center}
\end{table}

\begin{table}[!htbp]
\begin{center}
\begin{tabular}{lccccc}
\hline
\rule[-1mm]{0pt}{5mm}    $B=Z_{pole}$
  & $\kappa$
  & $\Lambda_{QCD}$
  & $a_1$
  & $a_2$
  & $\chi^2 /d.o.f.$  \\
\hline
\rule{0pt}{4mm}$16^3\times 128$ & $0.5335^{+23}_{-26}$ 
                 & $373^{+3}_{-3}$ 
                 & $0.0081^{+3}_{-3}$ 
                 & $2.36^{+2}_{-2}$  
                 & $1.65$ \\
\rule{0pt}{4mm}$16^3\times 256$ & $0.5217^{+14}_{-14}$ 
                 & $377^{+2}_{-2}$  
                 & $0.0082^{+3}_{-3}$ 
                 & $2.36^{+1}_{-1}$ 
                 & $1.61$ \\
\rule[-1mm]{0pt}{5mm}$18^3\times 256$ & $0.5300^{+17}_{-21}$ 
                 & $388^{+3}_{-3}$  
                 & $0.0085^{+6}_{-4}$ 
                 & $2.46^{+2}_{-3}$ 
                 & $1.20$ \\
\hline
\end{tabular}
\caption{Fits to all lattice data using $\alpha_{LN}(q^2)$ and $B=Z_{pole}$.
\label{dse2_b_pole}}
\end{center}
\end{table}

The running coupling at zero momentum can be measured from $\alpha_{LN}(q^2)$
\cite{sardenha, dublin}
and is related with the high momentum behaviour of the running coupling, 
$\alpha(0) ~ = ~ (4\pi/\beta_0) \, a_2 \,$.
The measured $\alpha(0)$ are reported in table \ref{alphtable}. 
Note that the values from
the $16^3 \times 128$ and $16^3 \times 256$ data agree within one standard
deviation. Furthermore, $\alpha(0)$ increases with $L_s$ and the 
fitted values reported
in table \ref{alphtable} are around
the original DSE estimation for $\alpha(0)=2.972$ \cite{lerche}.

\begin{table}[!htbp]
\begin{center}
\begin{tabular}{ccc}
\hline
\rule[-1mm]{0pt}{5mm}$\alpha(0)$ &  $ Z_{cut}$  & $Z_{pole}$  \\
\hline
\rule{0pt}{4mm} $16^3\times128$  &  $2.79^{+2}_{-1}$ & $2.70^{+2}_{-2}$\\
\rule{0pt}{4mm} $16^3\times256$  &  $2.77^{+1}_{-1}$ & $2.70^{+1}_{-1}$\\
\rule[-1mm]{0pt}{5mm} $18^3\times256$  &  $2.91^{+2}_{-3}$ & $2.81^{+2}_{-3}$\\
\hline
\end{tabular}
\caption{Values of $\alpha(0)$ measured from $\alpha_{LN}(q^2)$.\label{alphtable}}
\end{center}
\end{table}


\section{Gribov copies effects in the propagators}

The propagators are gauge dependent quantities. To compute them, one has 
to choose a gauge. For the Landau gauge on the lattice, the procedure 
consists in maximizing the functional,

\begin{equation}
   F_{U}[g] ~ = ~ C_{F}\sum_{x,\mu} \, \mbox{Re} \, \{ \, \mbox{Tr} \,
       [ g(x)U_{\mu}(x)g^{\dagger }(x+\hat{\mu}) ] \, \}  \label{f}
\end{equation}

It is well known that this functional has, in general, 
several maxima, the Gribov copies \cite{gribov}. 
To properly define a non-perturbative 
gauge fixing, one should 
choose the gauge transformation $g(x)$ that globally maximizes (\ref{f}). 
This is a global optimization problem and it is not easy 
to find the global maximum of (\ref{f}). In most studies of 
the gluon and ghost propagators one chooses a local maximum of 
(\ref{f}), hoping that the Gribov copy effects in the propagators are small.

On the lattice, several studies reported Gribov effects in the 
ghost propagator (see, for example, 
\cite{cucc97,stern05,madrid}, and the next section), 
but it is generally accepted that Gribov copies do not change significantly 
the gluon propagator. 

However, there are some studies reporting Gribov copy effects in the gluon 
propagator. In \cite{npb04}, the problem of the Gribov copies 
effects on the gluon propagator was studied, and some differences were found.
It was claimed a two to three $\sigma$ effect due to Gribov copies in the 
low momentum region. 
Furthermore, it was observed, by sorting the different Gribov 
copies according to the value of the gauge fixing functional, that 
the propagator for the lowest momenta behaves monotonically as a 
function of $<F_{U}>$: for zero momentum, the propagator increases with 
$<F_{U}>$, and for some small non-zero momenta the propagator decreases 
with $<F_{U}>$. There is also a study \cite{bbmpm05} where the 
authors claim that the enlargement of the gauge orbits allows for Gribov 
effects in the gluon propagator.

\begin{figure}[htbp]
\begin{center}
\psfrag{EIXOX}{{\small $\hat{n}_t$}}
\psfrag{EIXOY}{{\small $D_{ID}/D_{CEASD}$}}
\includegraphics[width=8cm]{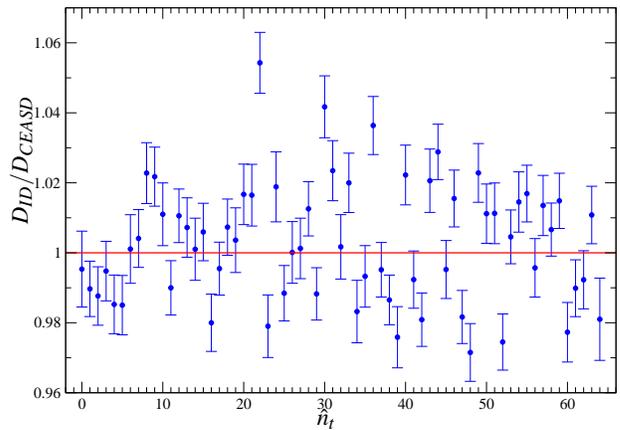}
\caption{Gribov copy effects in the gluon propagator computed from 
$16^3\times128$ configurations.}
\label{gribov.128}
\end{center}
\end{figure}

In what concerns asymmetric lattices, CEASD gauge fixing 
method \cite{ceasd} was applied to the $16^3\times 128$ configurations. 
This global optimization method for Landau gauge fixing combines a genetic 
algorithm with a local gauge fixing method, aiming to find the global 
maximum of (\ref{f}). The obtained gluon propagator was compared with 
the one computed from a local maximum of (\ref{f}); this local maximum 
is obtained using the local gauge fixing method described in \cite{davies}. 
In fig. \ref{gribov.128}, we can see the 
ratio between the two propagators.

Although there are many momenta for which the ratio is not compatible with 
one, we cannot conclude in favour of any systematic effect.

\section{Ghost propagator and running coupling constant}

\subsection{Ghost propagator}

For our smallest lattices ($10^3\times256$, $12^3\times256$, $16^3\times128$), 
we have computed the ghost propagator \cite{madrid},

\begin{equation}
G^{ab}(q)=-\delta^{ab} G(q^2)
\end{equation}

\noindent
 using both a point source method 
\cite{suman} (we averaged over 7 different point sources to get a better 
statistics), and a plane-wave source \cite{cucc97}. In both cases we used 
the preconditioned conjugate algorithm, as described in \cite{stern05}. 
The advantage of using a plane-wave source being that 
the statistical accuracy is much better, but we can only obtain one momentum 
component at a time. Using a point source method one can get all the momenta 
in once, but with larger statistical errors.

\begin{figure}[htbp]
\begin{center}
\vspace*{0.7cm}
\psfrag{EIXOX}{$q(GeV)$}
\psfrag{EIXOY}{ $Z_{ghost}(q^2)$}
\includegraphics[width=8cm]{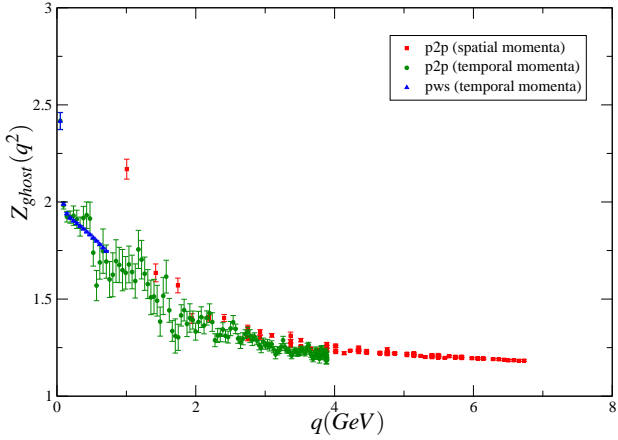}
\end{center}
\caption{Bare ghost dressing function for a $12^3\times256$ lattice; 
``p2p'' (``pws'') stands for the ghost components computed using a 
point (plane wave) source.}
\label{ghost_12.3.256}
\end{figure}

\begin{figure}[htbp]
\begin{center}
\vspace*{0.7cm}
\psfrag{EIXOX}{$q(GeV)$}
\psfrag{EIXOY} {$Z_{ghost}(q^2)$}
\includegraphics[width=8cm]{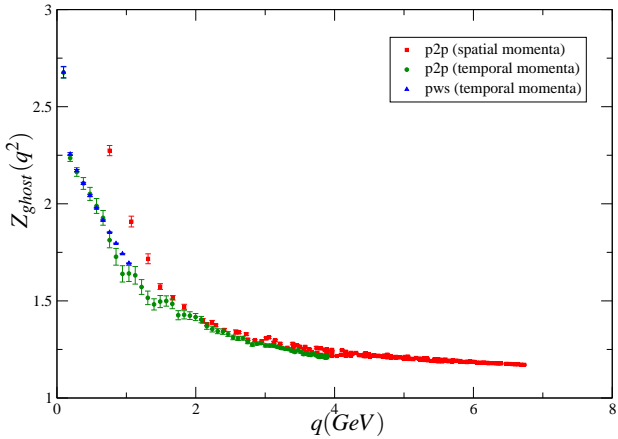}
\end{center}
\caption{Bare ghost dressing function for a $16^3\times128$ lattice.}
\label{ghost_16.3.128}
\end{figure}

In figures \ref{ghost_12.3.256} and \ref{ghost_16.3.128} it is shown the 
ghost dressing function $Z_{ghost}( q^2 )$ for $12^3\times256$ 
and $16^3\times128$ computed with both methods. For the $12^3\times256$, 
the point source data is not smooth for a large range of momenta. 
This is true also for $10^3\times256$ 
lattice , but this effect is significantly reduced for $16^3\times128$. 
In what concerns 
finite volume effects, we see differences, in the infrared, between pure 
temporal and pure spatial data, as in the gluon case.

\begin{figure}[htbp]
\begin{center}
\vspace*{0.7cm}
\psfrag{EIXOX}{$q(GeV)$}
\psfrag{EIXOY}{$Z_{ghost}(q^2)$}
\includegraphics[width=8cm]{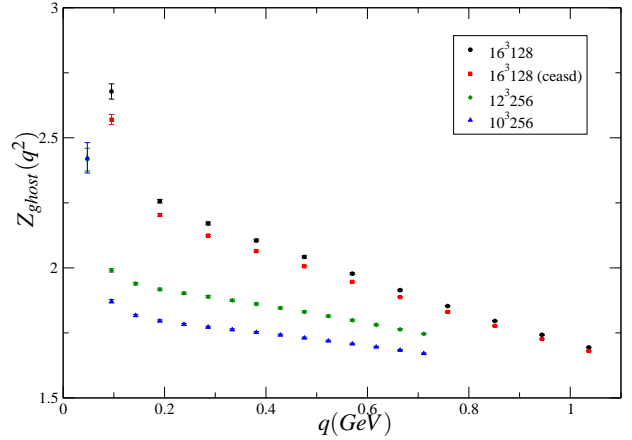}
\end{center}
\caption{Bare ghost dressing function for all lattices. The data were computed 
using a plane wave method.}
\label{ghost_cucc}
\end{figure}

In figure \ref{ghost_cucc} one can see the ghost dressing function only for 
the plane-wave data. As in the gluon case, we are able to evaluate the
 effect of Gribov 
copies on the lattice $16^3\times128$ by considering different gauge fixing 
methods. One can see clear effects of Gribov copies, as expected 
from other studies \cite{cucc97,stern05}. Also, we see finite volume effects 
if one compares propagators computed from lattices with different 
spatial sizes. 

In what concerns the infrared region, we were unable to fit a pure power law, 
even considering polinomial corrections \cite{madrid}. This negative 
result can be due either to the finite volume effects caused by the small 
spatial extension of our lattices, or to an insufficient number of 
points in the infrared region --- note that for the ghost propagator, the 
pure power law lacks validity well below 200 MeV (see figure \ref{dsepower}).

\subsection{Running coupling constant}

Using the gluon and ghost dressing functions, one can define a 
running coupling constant as

\begin{equation}
   \alpha_{S}(q^2) = \alpha(\mu^2) Z_{ghost}^2(q^2) Z_{gluon}(q^2).
\end{equation}

The DSE infrared analysis predicts a running coupling constant at zero momentum
different from zero, $\alpha_{S}(0)=2.972$ \cite{lerche}. On the other 
hand, the DSE solution on a torus \cite{dsetorus}, and results from 
lattice simulations \cite{stern05, furui1, furui2, madrid}, shows a 
decreasing coupling constant 
for small momenta. Using an asymmetric lattice allow us to study smaller 
momenta having in mind to provide, at least, a hint to this puzzle.

\begin{figure}[htbp]
\begin{center}
\vspace*{0.7cm}
\psfrag{EIXOX}{$q(GeV)$}
\psfrag{EIXOY}{$\alpha(q^2)/\alpha(\mu^2)$}
\includegraphics[width=8cm]{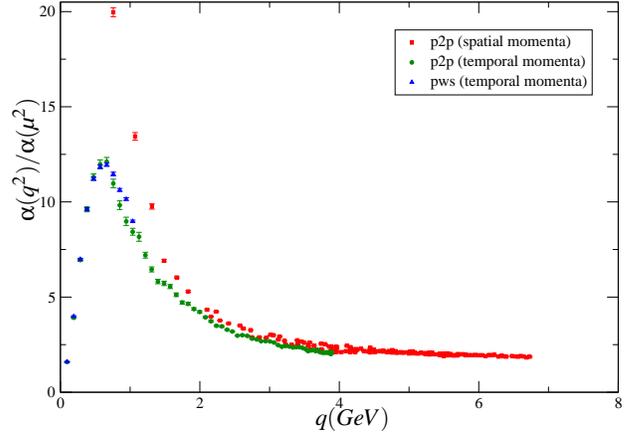}
\end{center}
\caption{Running coupling constant for a $16^3\times128$ lattice.}
\label{alpha.16.3.128}
\end{figure}

\begin{figure}[htbp]
\begin{center}
\vspace*{0.7cm}
\psfrag{EIXOX}{$q(GeV)$}
\psfrag{EIXOY}{$\alpha(q^2)/\alpha(\mu^2)$}
\includegraphics[width=8cm]{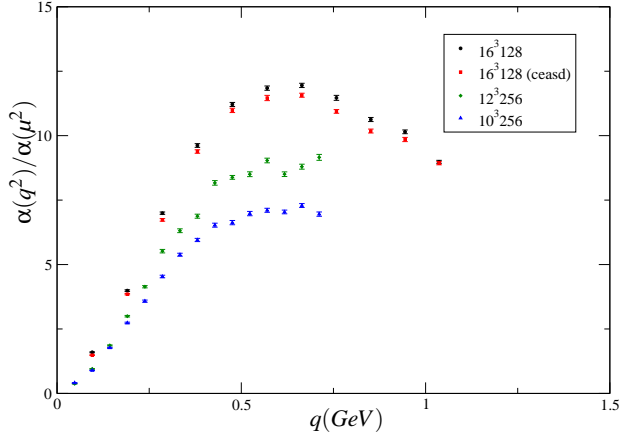}
\end{center}
\caption{Running coupling constant for all lattices. The data were 
computed using a plane wave method.}
\label{alpha.cucc}
\end{figure}

Again, our lattice data shows finite volume effects, if one compares pure 
temporal and pure spatial momenta, see figure \ref{alpha.16.3.128}.
Comparing the results for all available lattices (plane-wave source), 
see figure \ref{alpha.cucc}, we can see, as in the ghost case, 
finite volume effects, and clear Gribov copies effects.

In what concerns the infrared behaviour, we tried to fit the lowest momenta 
to a pure power law, $(q^2)^{\kappa_{\alpha}}$. We concluded that this 
power law is only compatible with the data from $16^3\times128$ lattice, 
gauge fixed with CEASD method, giving $\kappa_{\alpha}\sim 0.688$, with 
$\chi^{2}/d.o.f.\sim 0.011$. The reader should be aware that it is also 
possible, in some cases, to fit the infrared data to 
$\alpha(0) (1+ aq^2+\ldots)$ and get 
a $\alpha(0)\neq 0$. Therefore, we can not give a definitive answer about the 
behaviour of the running coupling constant for $q=0$. Note, however, that 
$\alpha_S (q^2)$ for the smallest momenta, seems to increase as 
a function of the volume.

\vspace*{1.5cm}
\section{Conclusions}

In this work asymmetric lattices were used to study the infrared behaviour of 
QCD propagators.

In what concerns the gluon propagator, the lattice data is well described by 
a pure power law $(q^2)^{2\kappa}$ for momenta below 150 MeV. The value of the 
infrared exponent increases with the volume, so  our $\kappa$'s can be read as 
lower bounds of the infinite volume figure. Extrapolating the gluon propagator 
to the infinite volume limit, assuming a sufficient temporal size of our 
lattices, one gets $\kappa\in [0.49,0.53]$. Unfortunately, we can not 
give a definitive 
answer about the behaviour of the gluon propagator at zero momentum. Note, 
however, that the lattice data favours the right hand side of the given 
interval, and that using other fitting forms and a larger range of momenta, 
one always get $\kappa>0.5$. 

In what concerns the ghost propagator, we were not able to extract a pure 
power law on the available results. We observed finite volume effects and 
clear Gribov copies effects on this propagator, in agreement with other studies.

Finally, we observed a decreasing running coupling for low momenta,
 in agreement with previous simulations.

\vspace*{1.0cm}
\section*{Acknowledgements}

This work was supported by FCT via grant SFRH/BD/10740/2002, and by project 
POCI/FP/63436/2005.





\end{document}